\documentclass[twocolumn,showpacs,preprintnumbers,amsmath,amssymb]{revtex4}
\usepackage{graphicx}% Include figure files
\usepackage{dcolumn}% Align table columns on decimal point
\usepackage{bm}% bold math
\newtheorem{Theo}{Theorem}
\newtheorem{Coro}{Corollary}
\begin{document}

%\preprint{APS/123-QED}

\title{Nonchaotic Stagnant Motion in a Marginal Quasiperiodic Gradient System\\}% Force line breaks with \\

\author{Takahito Mitsui}
\email{t.mitsui@aoni.waseda.jp}
\affiliation{%
Department of Applied Physics, Faculty of Science and Engineering,
Waseda University, Tokyo 169-8555, Japan
}%
%\altaffiliation[Also at ]{Department of Applied Physics, Faculty of Science and Engineering,
%Waseda University}%Lines break automatically or can be forced with \\

%\affiliation{

\date{\today}% It is always \today, today,

\begin{abstract}
A one-dimensional dynamical system with a marginal quasiperiodic gradient is presented
as a mathematical extension of a nonuniform oscillator.
The system exhibits a nonchaotic stagnant motion, which is reminiscent of intermittent chaos.
In fact, the density function of residence times near stagnation points
obeys an inverse-square law, due to a mechanism similar to type-I intermittency.
However, unlike intermittent chaos, in which the alternation between long stagnant phases and
rapid moving phases occurs in a random manner, here the alternation occurs in a quasiperiodic manner.
In particular, in case of a gradient with the golden ratio, the renewal of the largest residence time 
occurs at positions corresponding to the Fibonacci sequence.
Finally, the asymptotic long-time behavior, in the form of a nested logarithm, is theoretically derived.
Compared with the Pomeau-Manneville intermittency,
a significant difference in the relaxation property of the long-time average of the dynamical variable is found.
%Valid PACS numbers may be entered using the \verb+\pacs{#1}+ command.
\end{abstract}

\pacs{05.45.Ac, 05.45.Pq}% PACS, the Physics and Astronomy
\maketitle

\section{Introduction}
Temporal intermittency is the occurrence of a signal accompanied by
random alternation between long laminar phases and relatively short bursts,
which is widely observed in nonequilibrium dynamical systems.
For example, the Rayleigh-B\'{e}nard convection \cite{rf:Berg1980}, the Belousov-Zhabotinsky reaction \cite{rf:Pome1981},
and an rf-driven Josephson junction \cite{rf:Yeh1983} exhibit intermittent phenomena.
In the field of nonlinear physics, intermittency is tacitly understood as the occurrence of intermittent chaos.

Here we confine ourselves to the Pomeau-Manneville intermittency \cite{rf:Mann1980}. 
The onset of intermittent chaos is associated with a loss of stability of periodic motion,
which is classified into three types: type-I (saddle-node bifurcation),
type-II (Hopf bifurcation), and type-III (subharmonic bifurcations).
The Pomeau-Manneville (PM) system, $x_{n+1}=T(x_n)=x_{n}+ax_{n}^z\,(\mbox{mod}\,1)$ $(z\geq 1, a>0)$,
is a typical model of a Poincar\'{e} map for intermittent chaos \cite{rf:Mann1980}.
The system is called non-hyperbolic, since it has an indifferent fixed point $x=0$ with $T'(x)=1$,
in the neighborhood of which laminar motions are generated,
and the ergodic measure $\rho (x)$ localizes as $\rho (x)\propto x^{-z+1}$
\cite{rf:Mann1980,rf:Geis1984,rf:Gasp1988,rf:Aiza2000,rf:Akim2006}.

One of the important problems in such a non-hyperbolic system is the appearance of nonstationarity.
In a nonstationary regime where $z\geq 2$, the ergodic measure $\rho (x)$ cannot be normalized,
and the $f^{-\nu }\,(\nu \geq 1)$ fluctuation is generated.
Nonstationarity can be shown, for example, in the behavior of the renewal function $H(n)=E(N_n)$, 
which is the ensemble average of the number $N_n$ of chaotic bursts 
during the time interval $(0,n]$ \cite{rf:Gasp1988,rf:Akim2006,rf:Cox1962}.
Therefore, the renewal rate defined by $H(n)/n$ represents
the average occurrence probability of the bursts during $(0,n]$.
In stationary regimes, this probability does not depend on time, $H(n)/n\simeq 1/\tau$.
However, in the nonstationary regime $(z>2)$, it depends on time as $H(n)/n \propto n^{-\frac{z-2}{z-1}}$.
In particular, at the critical point $(z=2)$, it behaves as $H(n)/n \propto 1/\ln n$.
Thus, the renewal rate indicates the nonstationarity of the intermittent chaos.

Although many studies of intermittency have dealt with chaotic systems, 
intermittent dynamics are also observed in nonchaotic systems.
Intermittent strange nonchaotic attractors (SNAs) are generically created in quasiperiodically forced systems
through quasiperiodic saddle-node bifurcations \cite{rf:Pras1997,rf:Venk2000,rf:Kim2003}, 
through quasiperiodic subharmonic bifurcations \cite{rf:Venk1999,rf:Venk2000}, and
through several types of crisis \cite{rf:Witt1997},
where the scaling behavior is characteristic of type-I, 
type-III, and crisis-induced intermittency, respectively.
Unlike the analogy between the intermittencies in SNAs and those in chaotic systems,
to the best of our knowledge, the difference is not clear,
except for the original difference in the sign of the largest nontrivial Lyapunov exponent
\cite{rf:Pras1997,rf:Witt1997,rf:Venk1999,rf:Venk2000,rf:Kim2003}.

In this study, we present a marginal quasiperiodic gradient system (MQPGS) as a mathematical extension 
of a nonuniform oscillator $\dot x =1 -A\cos x$, which arises in several fields,
such as electronics, condensed-matter physics, mechanics, and biology
\cite{rf:Stro1994,rf:Winf2001},
and study the stagnant motion generated by the system.
The term ``nonchaotic stagnant motion'' in this paper represents anomalous dynamics
accompanied by long laminar phases and those interruptions {\it not} based on chaotic dynamics.
Stagnant motion in a MQPGS is related to the spatial quasiperiodicity of the gradient,
which differs from the temporal quasiperiodicity in intermittent SNAs.
The quasiperiodic gradient could be implemented in some Josephson junction circuits. 
Our main result is the theoretical derivation of the asymptotic divergence 
of the displacement in a MQPGS.
By comparing the asymptotic behavior in the MQPGS with the renewal rate in a PM system,
we clarify a significant difference in the relaxation property
of the long-time average of the dynamical variable. 
 
This paper is organized as follows: 
Section~II introduces the MQPGS and mentions some implementation methods,
Section~III presents the analysis of a two-period case of a MQPGS, 
and Section~IV gives the derivation of the density function of residence times near stagnation points.
Section~V clarifies the parameter dependence of the largest residence time from a number-theoretic point of view,
Section~VI gives the derivation of the asymptotic behavior of the MQPGS, 
and the last section presents a comparison between the nonchaotic stagnant motion of MQPGS
and the intermittent chaos of the PM system from the viewpoint of asymptotic behavior.

\section{Marginal Quasiperiodic Gradient System}
A nonuniform oscillator \cite{rf:Stro1994} is described by the equation
\begin{equation}
\dot x=1 -A\cos x \label{eqn:nonuniform}
\end{equation}
in the time scale normalizing the phase-averaged angular frequency.
Here $A$ is the control parameter, and $A=0$ corresponds to the uniform oscillator.
The nonuniform oscillator has a phase-drift ($A<1$) or phase-locked state ($A>1$)
as a result of saddle-node bifurcation.
There are many oscillatory phenomena explained by Eq.~(\ref{eqn:nonuniform}),
such as those in oscillating neurons, firefly flashing, a Josephson junction
and an overdamped pendulum driven by a constant torque \cite{rf:Stro1994,rf:Winf2001}.
  
As a mathematical extension, it is natural to ask what happens when the right-hand side of Eq.~(\ref{eqn:nonuniform}) becomes quasiperiodic. 
Thus, we introduce an MQPGS described by
a one-dimensional ordinary differential equation
\begin{equation}
\frac{dx(t)}{dt}=1-A_1\cos (k_1x+\delta _1)-A_2\cos (k_2x+\delta _2), \label{eqn:MQPGS} 
\end{equation}
where $A_i>0,\,k_i,\,\delta _i$ ($i=1,2$) are parameters, 
$k_1/k_2$ is irrational, and $A_1+A_2=1$.
We are also interested in the rational system ($k_1/k_2$ is rational), 
since the MQPGS is indistinguishable from well-approximated rational systems 
in a finite time and with finite resolution.

\begin{figure}[]
\includegraphics[scale=0.25, angle=0]{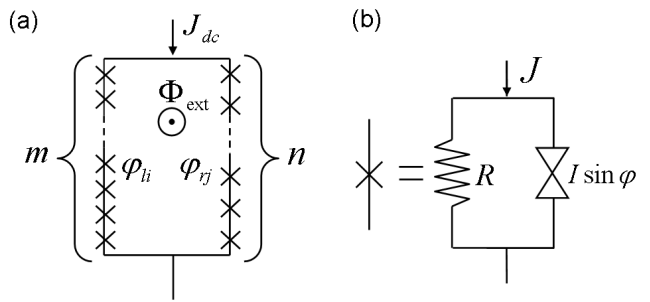}
\caption{(a) Asymmetric multijunction SQUID for an implementation of MQPGS. 
(b) Resistively shunted junction model of a single Josephson junction in overdamped limit.
}
\end{figure}
The MQPGS could be implemented in asymmetric multijunction superconducting quantum interference devices (SQUID)
modeled after 3JJ SQUID ratchet proposed by Zapata et al. \cite{rf:Zapa1996}.   
Figure~1 shows the schematic representation of an asymmetric multijunction SQUID threaded by a flux $\Phi _{\text{ext}}$, 
where the left and right branches contain $m$ and $n$ identical Josephson junctions, respectively.
The SQUID is driven by a dc current $J_{dc}$, which splits into two branch currents, $J_l$ (left) and $J_r$ (right).
We assume that the junctions are described by the resistively shunted junction (RSJ) model \cite{rf:Stew1968,rf:MuCu1968,rf:Baro1982}
with resistance $R$ and critical current $I$. 
Furthermore, we consider the overdamped limit, $(2e/\hbar )IR^2C\ll 1$, 
where the junction capacitance $C$ is negligible (see Fig.~1(b)), and also the Nyquist noise current is neglected.
Then, the phase $\phi _{li}$, across the single junction $i$ on the left branch, obeys the following equation:
\begin{equation}
\frac{\hbar }{2eR}\dot \phi _{li}+I_l\sin (\phi _{li})=J_l
\label{eqn:SQUID}
\end{equation}
On the right arm, the phase $\phi _{rj}$ of the junction $j$ obeys the equation obtained by
replacing labels $l$ and $i$ in Eq.~(\ref{eqn:SQUID}) with $r$ and $j$, respectively.
If identical initial conditions are assumed, the only solutions for each junction 
are phase-locked states, $\phi _{l1}=\phi _{l2}=\cdots =\phi _{lm}=\phi _l$ and 
$\phi _{r1}=\phi _{r2}=\cdots =\phi _{rn}=\phi _r$.
In the limit where the total loop inductance $L$ is negligible, $|LI|\ll \Phi _0$, 
the total flux is approximately equal to the external flux $\Phi _{\text{ext}}$, 
where $\Phi _0\equiv \hbar /2e$ is the flux quantum.
Then, the integration of the gauge invariant phase around the loop yields
$m\phi _l -n\phi _r=-2\pi \Phi _{\text{ext}}/\Phi _0+2\pi s$ $(s\in \mathbb{N})$.
Consequently, the Kirchhoff's current law $J_{dc}=J_l+J_r$ is given by
\begin{eqnarray*}
\frac{\hbar }{2eR}\left( 1+\frac{m}{n}\right) \dot \phi _l
&=&J_{dc}-I_l\sin (\phi _l)-I_r \\
&&\times \sin \left( \frac{m}{n}\phi _l+2\pi \frac{\Phi _{\text{ext}}/\Phi _0-s}{n}\right) ,
\end{eqnarray*}
which approximately describes the MQPGS with $k_2/k_1\approx m/n$, under the condition of 
$J_{dc}=I_l+I_r$, if all the variables are transformed properly.
Note that the better approximation of $k_2/k_1\approx m/n$ yields the better implementation.

Also, the MQPGS could be implemented in a dc-driven circuit consisting of overdamped Josephson junctions coupled by ideal transformers.
However, we omit a detailed description, since this is not the purpose of this paper.  

\section{Nonchaotic Stagnant Motion}
In the following sections, we analyze dynamics of MQPGS, focusing on a typical case:
\begin{equation}
\frac{dx(t)}{dt}=1-\frac{1}{2}\cos (2\pi x)-\frac{1}{2}\cos (2\pi kx), \label{eqn:model}
\end{equation}
where $k$ is a control parameter, and $x(0)=x_0$.
\begin{table}
\caption{Parameters for lines in Fig.~2}
\begin{ruledtabular}
\begin{tabular}{cccc}
  Line  & $k$ & Continued fraction \footnotemark[1] & Approximation \\ \hline
  (a)& $\frac{102334155}{165580141}$ & $[0;\,\bar{1}^{40}]$ & 0.618033988749894 \\ 
  (b)& $\frac{148099316}{239629871}$ & $[0;\,\bar{1}^{14},\,2,\,\bar{1}^{25}]$ & 0.618033617353155 \\ 
  (c)& $\frac{514220604}{832027711}$ & $[0;\,\bar{1}^{14},\,10,\,\bar{1}^{25}]$ & 0.618033260658830 \\ 
  (d)& $\frac{377}{610}$ & $[0;\,\bar{1}^{14}]$ & 0.618032786885245 
  \end{tabular}
\end{ruledtabular}
\footnotetext[1]{For example, the notation $[0;\,\bar{1}^{14},\,2,\,\bar{1}^{25}]$
means a finite continued fraction $\left[ q_0;\,q_1,\,q_2,\,\,...,q_{40}\right]$ 
with partial quotients, $q_0=0$, $q_1=q_2=\cdots =q_{14}=1$,
$q_{15}=2$, and $q_{16}=\cdots =q_{40}=1$.}
\end{table}
\begin{figure}[]
\includegraphics[scale=0.55, angle=0]{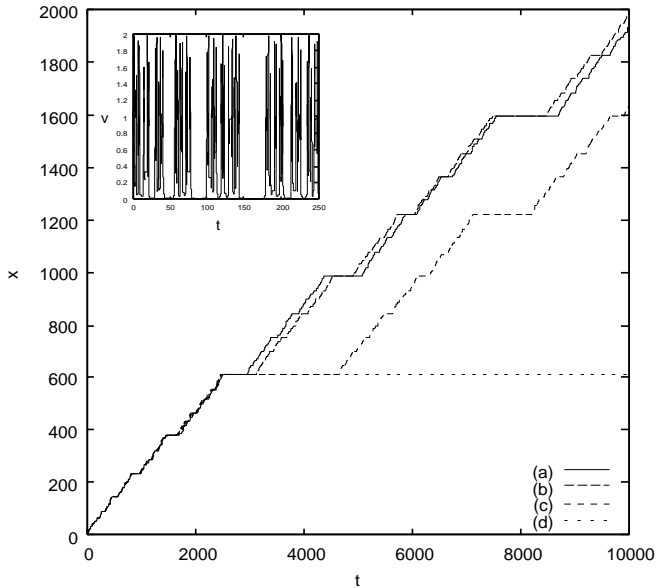}
\caption{Displacement trajectories $x(t)$ starting from the same initial condition $x_0=1/2$
for various values of parameter $k$ 
near the inverse golden ratio $\varphi ^{-1}=\frac{\sqrt{5}-1}{2}$.
$k$ values for each line are given in Table~1.
The inset shows the velocity $v(t)$ for $k=\varphi ^{-1}$.}
\end{figure}
The displacement $x(t)$ never decreases
but frequently slows down due to the quasiperiodic gradient.
Figure~2 shows the trajectories of the displacement $x(t)$ starting from the same initial condition $x_0=1/2$
for various values of $k$ near the inverse golden ratio $\varphi ^{-1}=\frac{\sqrt{5} -1}{2}.$
In the case where $k$ is a rational number, $p/q$, with
co-prime integers $p$ and $q$,
the displacement $x(t)$ finally goes to one of the equilibrium states $x=qn\,(n\in \mathbb{Z})$.
Actually, the line (d) in Fig.~2 is for $k=377/610$, 
which has an equilibrium point at $x=610$.
If the parameter $k$ is $p'/q'$, which differs slightly from $p/q$, and $q'>q$,
the displacement $x(t)$ takes a long time to pass
through the vicinities of equilibrium points for $k=p/q$. We call these stagnant phases. 
Lines~(a)-(c) in Fig.~2 represent the displacements for several $k$-values slightly different from $337/610$.
We observe the alternation between long stagnant phases
and rapid moving phases, which is reminiscent of intermittent chaos (see inset of Fig.~2).
When $k$ is irrational, the equilibrium points vanish except for the origin, 
but the motion stagnates in the vicinity of $x=qn$
if the value of $k$ is close to $p/q$. 
Note that every rational $k$ is a bifurcation point;
i.e., the system is structurally unstable with respect to $k$.

When the motion stagnates, the factor $\cos (2\pi x)$ must approach unity.
Thus, the coordinates of the stagnation points are restricted to the vicinities of integers $n$.
Hence, we consider the relative dynamics of the variable $y=x-n\,(-\frac{1}{2}\leq y\le \frac{1}{2})$
within each cells $I_n\equiv \left[ n-\frac{1}{2},n+\frac{1}{2}\right]$,
and introduce new parameters:
\begin{eqnarray}
\gamma _n&\equiv & \left \{
\begin{array}{l}
kn\,\text{(mod 1)}\,\,\,\,\,\,\,\,\,\,\,\,\,\,\text{if}\,\,kn\,\text{(mod 1)}\le 1/2,\\ \nonumber
kn\,\text{(mod 1)}-1\,\,\,\,\text{if}\,\,kn\,\text{(mod 1)}>1/2, \nonumber
\end{array}
\right. \\
\varepsilon _n &\equiv & |\gamma _n|=\mbox{min}[kn\,\text{(mod 1)},\,1-kn\,\text{(mod 1)}].\nonumber
\end{eqnarray}
Then, the dynamics in the {\it n}-th cell is written as
\begin{equation}
\frac{dy(t)}{dt} = 1-\frac{1}{2}\cos (2\pi y)-\frac{1}{2}\cos (2\pi (\gamma _n+ky)).
\label{eqn:modeleps}
\end{equation}
The parameters $\gamma _n$ and $\varepsilon _n$ are uniquely determined
for each $n$ in $-\frac{1}{2}< \gamma _n\leq \frac{1}{2}$
and in $0\leq \varepsilon _n\leq \frac{1}{2}$, respectively.
They relates to the amount of the stagnation in each cell.

\section{Density Function of Residence times}
In the following, let $k$ be irrational, and the initial condition be $x_0=1/2$.
Then, the displacement $x(t)$ visits every cell $I_n\,\,(n\geq 1)$ only once in the course of time.
Let us define the residence time $T _n$ as the period
for which the object stays in the $n$-th cell $I_n$ as
$T_{n}= \int_{-\frac{1}{2}}^{\frac{1}{2}} \frac{dy}{1-\frac{1}{2}\cos (2\pi y)-\frac{1}{2}\cos (2\pi (\gamma _n+ky))}.$
Due to the symmetry of the integral interval, $\gamma _n$ in the integral can be replaced with $\varepsilon _n$. 
Thus, the residence time $T_n$ is written as a function of $\varepsilon _n$,
\begin{equation}
T_{n}= \int_{-\frac{1}{2}}^{\frac{1}{2}} \frac{dy}{1-\frac{1}{2}\cos (2\pi y)-\frac{1}{2}\cos (2\pi (\varepsilon _n+ky))}.
\label{eqn:defT}
\end{equation}
To estimate the residence time, 
consider the narrow $y$-region for each cell,
in which $\cos(\pi y)$ and $\cos (\pi (\gamma _n+ky))$
approach unity simultaneously, and are 
approximated by the second-order Taylor expansion around zero for each phase.
When $k<1$ is satisfied, at most only one such region is present within each cell,
and is expressed by parameter $r$ near unity as follows:
\begin{eqnarray*}
&&I_{n,r} = \{\,y\,|\cos (2\pi y)>r\,\wedge  \,\cos (2\pi (\gamma _n+ky))>r\}, \\
&&= \left( \max \left[ -\delta (r),-(\delta (r)+\gamma _n)/k \right] ,\min \left[ \delta (r),(\delta (r)-\gamma _n)/k \right] \right),
\end{eqnarray*}
where $\delta (r)=\frac{1}{2\pi} \arccos (r)$.
For $k>1$, Eq.~(\ref{eqn:modeleps}) is reduced to a similar form for $0<k'<1$
using scale transformation
$k'=1/k$, $x'=kx$, and $t'=kt$.
Therefore, we limit our discussion to the case of $0<k<1$.
In each region $I_{n,r}$, Eq.~(\ref{eqn:model}) is approximated by the second-order Taylor expansion, 
\begin{equation}
\frac{dy}{dt}\simeq \pi ^2(1+k^2)\left( y+\frac{k}{1+k^2}\gamma _n \right) ^2
+\frac{\pi ^2 \varepsilon _n ^2}{1+k^2}.
\label{eqn:modeltaylor}
\end{equation}
The second term of Eq.~(\ref{eqn:modeltaylor}) gives the minimum velocity in each cell.
The residence time $T_{n,r}$ in the region $I_{n,r}$ is given
by the integration of Eq.~(\ref{eqn:modeltaylor}),
\begin{widetext}
\begin{eqnarray}
T_{n,r} &\simeq & \frac{1}{\pi ^2 (1+k^2)} \Biggl[ \frac{1+k^2}{\varepsilon _n} \arctan \left[ \frac{(1+k^2)y}{\varepsilon _n} \right] \Biggr]
             _{\max \left[ -\delta (r),-(\delta (r)+\gamma _n)/k \right]+\frac{k\gamma _n }{1+k^2}}
             ^{\min \left[ \delta (r),(\delta (r)-\gamma _n)/k \right]+\frac{k\gamma _n }{1+k^2}},\nonumber \\ 
             &\rightarrow & \frac{1}{\pi \varepsilon _n}\,\,\,\,(\varepsilon _n\rightarrow 0).
             \label{eqn:residencetime}
\end{eqnarray}
\end{widetext}
In the complementary regions $I_n\setminus I_{n,r}$, the residence times are shorter than $1/(1-r)$, since $\dot y>1-r$.
Therefore, the residence time $T_n$ is dominated by $T_{n,r}$ for small $\varepsilon _n$, ${\it i.e.}$,
\begin{equation}
T_n \sim \frac{1}{\pi \varepsilon _n}.
\label{eqn:Tsimeps}
\end{equation}
This is the same as the universal scaling of type-I intermittency with an index of $-1/2$,
since $T_n\sim (\pi ^2\varepsilon _n^2)^{-1/2}$ when the channel width is $\pi ^2 \varepsilon _n ^2/(1+k^2)$.

After sufficient displacement $x(t)$, the density function of observed $\varepsilon _n$ converges to 
a uniform density $F(\varepsilon _n)=2$ on the interval $\left( 0,\frac{1}{2}\right)$,
since the values of $kn \pmod{1}\,\,(n\in \mathbb{N})$ are distributed uniformly on the torus interval $(0,1)$ \cite{rf:Arno1968}.
As a result, the density function $P(T)$ of residence times is obtained by $P(T) = -F(\varepsilon )\frac{d\varepsilon }{dT}$
and shows an inverse-square law when the residence time $T$ is sufficiently large, 
\begin{equation}
P(T) \rightarrow  \frac{2}{\pi T^2}\,\,\,\,(T \rightarrow \infty ).
     \label{eqn:densitydist}
\end{equation}
Note that this density function does not have any finite moments.

\begin{figure}[tb]
\includegraphics[scale=0.55, angle=-90]{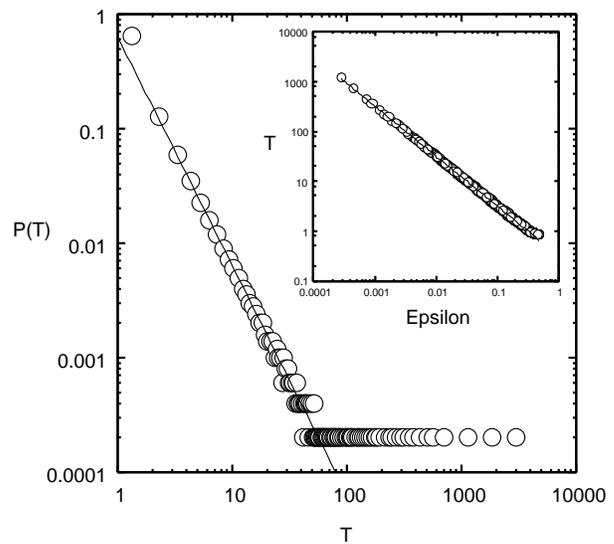}
\caption{The density function $P(T)$ of residence times for $k=\varphi ^{-1}$.
Residence times from $n=1$ to $5000$ are used. The bin width is set to 1.
The solid line represents a theoretical estimate obtained from Eq.~(\ref{eqn:densitydist}).
Deviations are observed for large $T$, where the statistics are poor.
The inset shows the residence times $T_n$ as a function of $\varepsilon _n$.
The residence times $T_n$ from $n=1$ to 2000 are plotted for $k=\varphi ^{-1}$.
The solid line is the theoretical estimate from Eq.~(\ref{eqn:Tsimeps}).
Deviations from the theoretical estimate $|T_n-\frac{1}{\pi \varepsilon _n}|$
converge to about $0.063\pm 0.001$ when $\varepsilon _n< 0.001$.}
\end{figure} 
The numerical results agree with Eqs.~(\ref{eqn:Tsimeps}) and (\ref{eqn:densitydist}) in their asymptotic regions,
as shown in Fig.~3.
It is significant that the asymptotic behavior of the density function $P(T)$
is independent of parameter $k$, provided that $k$ is irrational.
Nevertheless, the observation time for obtaining the universal density function depends substantially on $k$.

%%%%%%%%%%%%%%%%%%%%%%%%%%%%%%%%%%%%%%%%%%%%%%%%%%%%%%%%%%%%%%%%%%%%%%%%%%%%%%%%%%%%%%%%%%%%%%%%%%%%%%%%%%%%%%%%%%%%%%%%%%%%%%%%%%%%%%%%%%%%%%%%%%%%%
\section{Renewal Process of the largest residence time}
Let us consider that the displacement $x(t)$ has passed
through the finite space interval $\left[ x_0,\,n_l+\frac{1}{2}\right)$,
where $n_l$ is the index of the latest cell $I_{n_l}$ that the displacement passed through.
Then, the largest residence time during the passage up to the cell $I_{n_l}$ is defined by
\begin{equation}
T_{n \leq n_l} ^{\max}=T_{n< n_l+1}^{\max}=\max \{ T_n\,|\,1\leq n\leq n_l\} . \nonumber
\end{equation}
When the number $n_l$ of passed cells increases,
the largest residence time $T_{n \leq n_l} ^{\max}$ is renewed.
This section clarifies the positions of cells where the renewal occurs.

The parameter $k$ can be expressed by the continued fraction representation as 
$k=\left[ q_0;\,q_1,\,q_2,\,\cdots ,q_{i-1},\,\theta _{i}\right] ,\,\,\,\mbox{where}\,\,\,
q_0\in \mathbb{Z},\,q_i\in \mathbb{N},\,\theta _i\in \mathbb{R},\,\,\mbox{and}\,\,\theta _i>1.$
$\{ q_i\}$  are called partial quotients, defined by the formula $q_i=\lfloor \theta _i\rfloor$.
$\theta _i$ is the $i$-th complete quotient, generated by the
recursion formula $\theta _i=q _i+\frac{1}{\theta _{i+1}}$.
Replacing $\theta _i$ in the continued fraction with $q_i$,
we get the $i$-th convergent of $k$, {\it i.e.},
$P_i/Q_i=[q_0;\,q_1,\,q_2,\,\cdots ,q_{i-1},\,q_i].$
These successive convergents are generated by the following recursion relations: \cite{rf:Rock1992}
\begin{eqnarray}
P_i &=& q_iP_{i-1}+P_{i-2},\,\,\,\,P_{-2}=0,\,\,P_{-1}=1,\nonumber \\
Q_i &=& q_iQ_{i-1}+Q_{i-2},\,\,\,\,Q_{-2}=1,\,\,Q_{-1}=0. \label{eqn:recursive}
\end{eqnarray}
Note that the sequences $\{P_i\}$ and $\{Q_i\}$ increase monotonically. 

Let $m$ be the nearest integer to $kn$.
Then, $\varepsilon _n$ is given by $\varepsilon _n=|kn-m|$.
Using the above parameters, we can rewrite $\varepsilon _n$ for $n=Q_i\,(i\geq 1)$ as
\begin{equation}
\varepsilon _{Q_i}=\left| kQ_i-P_i\right| =\frac{1}{\theta_{i+1} Q_i+Q_{i-1}}\,\,\,\,(i\geq 1),
\label{eqn:eps}
\end{equation}
which is derived in Appendix A. 

The following theorem of Lagrange states the order relations
in the sequence $\{ \varepsilon _n\}$ \cite{rf:Rock1992}.\\

\begin{Theo}
Let rational $m/n$ be different from either $P_i/Q_i$ or $P_{i+1}/Q_{i+1}$ with $1\leq n\leq Q_{i+1}$. Then,
$$|nk-m|>\left| kQ_i-P_i\right| >\left| kQ_{i+1}-P_{i+1}\right|, $$
i.e.,
\begin{eqnarray*}
&\varepsilon _j> \varepsilon _{Q_i} >\varepsilon _{Q_{i+1}}&\\
&\text{for any integer j satisfying}\,\,\,1\leq j<Q_{i+1}\,\,\,and\,\,\,j\neq Q_i.&
\end{eqnarray*}
\label{Theo1}
\end{Theo}
The proof is given in Appendix~B.\\

\noindent Since the residence times $T_n$ are determined by the parameters $\varepsilon _n$,
the next corollary follows for the residence times.\\

\begin{Coro}
Let $1\leq j<Q_{i+1}\,\,\,and\,\,\,j\neq Q_i.$ 
\begin{eqnarray*}
&& \varepsilon _j> \varepsilon _{Q_i} >\varepsilon _{Q_{i+1}},\\
\Rightarrow \,\,&& T_j< T _{Q_i} <T _{Q_{i+1}}.
\end{eqnarray*}
\label{Theo1}
\end{Coro}

\noindent Proof. The order relations in the sequence $\{ \varepsilon _n\}$ are obtained from Theorem~1.
The residence time $T_n$ is a monotonically decreasing function of $\varepsilon _n$,
since $\frac{dT_n}{d\varepsilon _n} <0$, as proven in Appendix C. 
Hence, the inequalities in residence times $\{T_n\}$ hold.
\hfill $\square$\\

Corollary~1 states that
the renewal of the largest residence time $T_{n \leq n_l}^{\max}$
occurs when the displacement $x(t)$ passes through cell $I_{Q_i}\,(i\geq 1)$, which includes
the stagnation points $x\approx Q_i$.
Therefore, the largest residence time is given simply by
\begin{equation}
T_{n<Q_{i+1}}^{\max}=T_{Q_i}\simeq \frac{\theta_{i+1} Q_i+Q_{i-1}}{\pi }.
\label{eqn:Tmax}
\end{equation}
Hence, if we know the convergent series of parameter $k$, 
the positions of cells where the renewal occurs are completely determined.
For example, for $k=\varphi ^{-1}$,
the stagnation points generating the largest residence time 
are determined by Eq.~(\ref{eqn:recursive}) with $q_0=0$ and $q_i=1\,\,\,\,(i\geq 1)$,
and given by $Q_i=\frac{\varphi ^{i+1}-(-\varphi ^{-1})^{i+1}}{\sqrt{5}}$
{\it i.e.}, the {\it Fibonacci sequence} (see Line~(a) in Fig.~2).
%%%%%%%%%%%%%%%%%%%%%%%%%%%%%%%%%%%%%%%%%%%%%%%%%%%%%%%%%%%%%%%%%%%%%%%%%%%%%%%%%%%%%%%%%%%%%%%%%%%%%%%%%%%%%%%%%%%%%%%%%%%%%%%%%%%%%%%%%%%%%%%%%%%%

\section{Asymptotic behavior of the displacement}
According to the renewal of the largest residence time,
the finite-time average of the velocity $\bar{v}(x,t)=x(t)/t$ is expected to decrease gradually as time $t$ grows,
although it stays positive. In this section, the asymptotic behavior of the displacement $x(t)$ is investigated in detail.

\subsection{Asymptotic estimate of the ratio $t/x(t)$}
The fluctuation of the ratio $t/x(t)$ characterizes the deviation from the linear increase in the displacement.
Let us define the ratio by $r(x)=t/x(t)$ as a function of the variable $x$ under the fixed initial condition $x_0=1/2$.

It can be proven that the ratio $r(x)$ satisfies the following inequalities for large integers $Q_i$:
\begin{eqnarray}
r\left( Q_i+\frac{1}{2}\right) \gtrsim
r\left( n+\frac{1}{2}\right) \gtrsim r\left( Q_{i+1}-\frac{1}{2}\right) \label{eqn:t/xmidineq}\\
\mbox{for}\,\,\,n=Q_i,\,Q_i+1,\,\cdots ,Q_{i+1}-1.\nonumber
\end{eqnarray}
The proof is given in Appendix~D by assuming the ergodic property of the sequence $\{\varepsilon _n\}$.
Each term in Eq.~(\ref{eqn:t/xmidineq}) is generally given by $r\left( n+\frac{1}{2}\right) \simeq \frac{1}{n}\sum _{j=1}^{n}T_j$
for large integers $n$, which is the arithmetical average of the residence times up to the $n$-th cell. 
Assume that the dynamics of the sequence $\{ \varepsilon _j\}_{j=1}^{Q_i-1}$ is approximately ergodic 
over the interval $[\varepsilon _{Q_{i-1}},\,\varepsilon ^{\max}]$ for large $Q_i$
where $\varepsilon ^{\max}$ denotes the maximum of $\varepsilon _j\,\,(j=1,\,\cdots ,\,Q_i-1)$,
and the minimum becomes $\varepsilon _{Q_{i-1}}$ from Theorem~1.
Then, the arithmetical average can be
replaced with the average calculated by the density function $P(T)$, whose asymptotic form is given by Eq.~(\ref{eqn:densitydist}).
Therefore, the ratio $r\left( Q_{i}+\frac{1}{2}\right)$ in the left part of Eq.~(\ref{eqn:t/xmidineq}) is approximated for large $Q_i$ as follows:
\begin{eqnarray}
r\left( Q_{i}+\frac{1}{2}\right)
               &\simeq & \frac{1}{Q_i-1}\sum _{j=1}^{Q_i-1}T_j+\frac{T_{Q_i}}{Q_i}, \nonumber \\
               &\simeq & \int^{T_{Q_{i-1}}}_{T^{\min}}TP(T)dT+\frac{T_{Q_i}}{Q_i},\nonumber \\
               &= & \frac{2}{\pi}\ln T_{Q_{i-1}}+C_0(k)+\frac{T_{Q_i}}{Q_i},\nonumber 
\end{eqnarray}
where $T^{\min}$ denotes the minimum residence time corresponding to $\varepsilon ^{\max}$, 
and the term $C_0(k)$ is defined by $C_0(k)=\int^{T_{Q_{i-1}}}_{T^{\min}}TP(T)dT-\frac{2}{\pi}\ln T_{Q_{i-1}}$,
the value of which depends on the behavior of $P(T)$ in the non-asymptotic region.
It is numerically confirmed that $C_0(k)$ converges to a constant for large $Q_i$ (in particular $C_0(\varphi ^{-1})\approx 0.558$).
From Eq.~(\ref{eqn:Tmax}), the ratio $r\left( Q_{i}+\frac{1}{2}\right)$ can be written 
solely with information about the rational approximation of parameter $k$,
\begin{eqnarray}
r\left( Q_{i}+\frac{1}{2}\right)
               &\simeq & \frac{2}{\pi}\ln (Q_{i}+Q_{i-1}/\theta _{i+1})\nonumber \\ 
&&+\frac{\theta _{i+1}Q_i+Q_{i-1}}{\pi Q_i}+C_1(k), \label{eqn:t/xQ+1/2}
\end{eqnarray}
where $C_1(k)=C_0(k)-\frac{2}{\pi}\ln \pi.$
In the same manner, the ratio $r\left( Q_{i+1}-\frac{1}{2}\right)$ in the right hand side of Eq.~(\ref{eqn:t/xmidineq}) 
is given for large $Q_{i+1}$ by
\begin{equation}
r\left( Q_{i+1}-\frac{1}{2}\right) \simeq  \frac{2}{\pi}\ln (Q_{i+1}+Q_{i}/\theta _{i+2})+C_1(k). \label{eqn:C0}
\end{equation}
Using Eqs.~(\ref{eqn:t/xmidineq}), (\ref{eqn:t/xQ+1/2}), and (\ref{eqn:C0}),
we can estimate the ratio $r(x)$ as follows:
\begin{widetext}
\begin{eqnarray}
&\frac{2}{\pi}\ln Q_{i+1}+C_1(k) \lesssim
r\left( n+\frac{1}{2}\right) \lesssim 
\frac{2}{\pi}\ln Q_i+\frac{2}{\pi}\ln \left( 1+\frac{1}{q_i\theta _{i+1}}\right) +\frac{\theta _{i+1}+1/q_i}{\pi}+C_1(k)&
\nonumber \\
&\mbox{for}\,\,\,\,n=Q_i,\,Q_i+1,\,\cdots ,Q_{i+1}-1.& \label{eqn:estimate}
\end{eqnarray}
%\end{widetext}
When $x=n+\frac{1}{2}\,(n=Q_i,\,Q_i+1,\,\cdots ,Q_{i+1}-1)$, $x$ satisfies $Q_i<x<Q_{i+1}$, and Eq.~(\ref{eqn:estimate})
reduces to 
\begin{eqnarray}
&\frac{2}{\pi}\ln x+C_1(k) <
\frac{t}{x} < 
\frac{2}{\pi}\ln x +\frac{\theta _{i+1}}{\pi}+\frac{2\ln 2+1}{\pi}+C_1(k),& \nonumber \\
&\mbox{for}\,\,\,\, x\in S_i\equiv \left\{ n+\frac{1}{2} \Bigm| n=Q_i,\,Q_i+1,\ldots Q_{i+1}-1 \right\}\,\,\text{and its index}\,\,i,&\nonumber
\end{eqnarray}
where $1/q_i\leq 1$ and $1/(q_i\theta _{i+1})<1$ are used.
\end{widetext}
The ratio $t/x$ can be expressed as the sum of a logarithmic term of $x$ and some bounded function $f(x)$, 
\begin{eqnarray}
&\frac{t}{x}\simeq \alpha \ln x+f(x),&\label{eqn:t/xf(x)} \\
&C_1(k)<f(x)<\frac{\theta _{i+1}}{\pi}+\frac{2\ln 2+1}{\pi}+C_1(k)& \label{eqn:f(x)}\\
&\mbox{for}\,\,\,\, x\in S_i\equiv \left\{ n+\frac{1}{2} \Bigm| n=Q_i,\,Q_i+1,\ldots Q_{i+1}-1 \right\}& \nonumber
\end{eqnarray}
and its index $i$, where $\alpha =2/\pi$.
The above estimates for the ratio $r(x)$ from (\ref{eqn:t/xQ+1/2}) to (\ref{eqn:f(x)})
are in good agreement with the numerical results.
Figure~4 shows the ratios $r(x)$ vs. $x$ for $k=\varphi ^{-1}$ and $k=1/e$ ($e=$ natural logarithm),
and the lower bound for $k=\varphi ^{-1}$ given by $\alpha \ln x+C_1(\varphi ^{-1})$.
Lower bounds for each ratio are nearly identical, since $C_1(\varphi ^{-1})\simeq C_1(1/e)$.
On the other hand, upper bounds vary depending on the parameter $k$.

If we know the value of the ($i$+1)-th complete quotient $\theta _{i+1}$ and the fraction $Q_i/Q_{i-1}$,
the lower and upper bounds for the ratio $r(x)=t/x$ for $x\in S_i$ are estimated
more accurately than by Eqs.~(\ref{eqn:t/xf(x)}) and (\ref{eqn:f(x)}).
For $k=\varphi ^{-1}$, relations $\theta _{i+1}=\varphi$ and $Q_i/Q_{i-1}\simeq \varphi$ are available for large $i$.
Letting $r_L(x)$ and $r_U(x)$ denote the lower and upper bounds, respectively,
the ratio $r(x)$ is estimated as follows:
\begin{eqnarray}
&&r_L(x)\lesssim r(x)\lesssim r_U(x)\,\,\,\,\mbox{for}\,\,\,\,k=\varphi ^{-1}, \nonumber \\
&&r_L\left( x\right) = \alpha \ln x+f_L, \label{eqn:t/xphi1} \\
&&r_U\left( x\right) = \alpha \ln x+f_L+(\varphi +\varphi ^{-1})/\pi \label{eqn:t/xphi2}
\end{eqnarray}
where $f_L =\alpha \ln \left( 1+\frac{1}{\varphi ^2}\right) +C_1(\varphi ^{-1})$.
Note that $r_L(x)$ and $r_U(x)$ are independent of the index $i.$ 

The inset of Fig.~4 shows that the ratio $r(x)$ for $k=\varphi ^{-1}$ has
zig-zag structures, which are similar to each other and appear at equal intervals in the logarithmic scale of $x$.
The large peaks correspond to the stagnation points generating the largest residence time,
which are distributed as the Fibonacci sequence,
$Q_i\simeq \frac{\varphi ^{i+1}}{\sqrt{5}}$ for large $i$.
\begin{figure}[]
\includegraphics[scale=0.425, angle=-90]{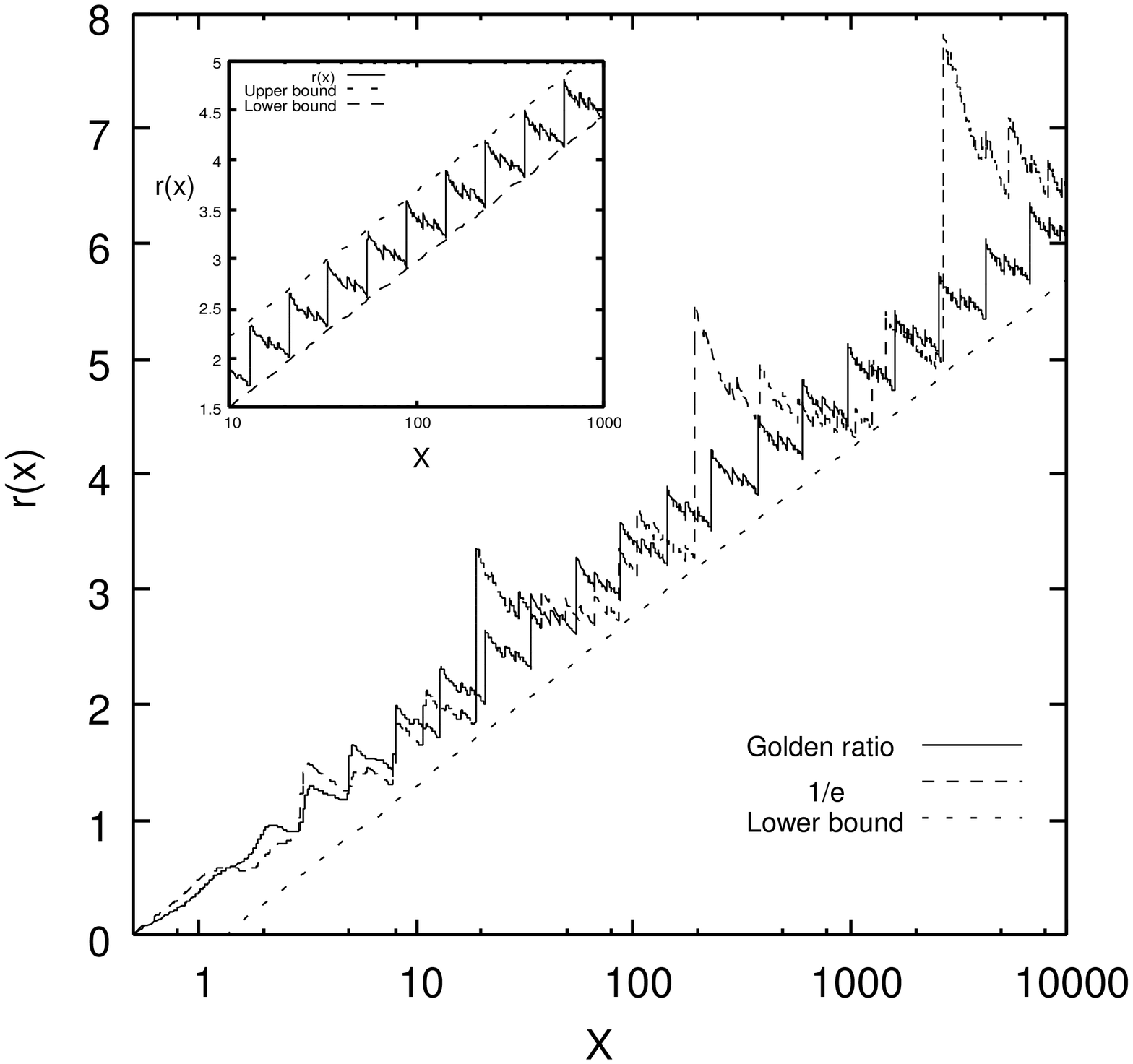}
\caption{Ratios $r(x)$ vs. displacement $x$ for $k=\varphi ^{-1}$ and $k=1/e$ ($e=$ natural logarithm).
The dashed line is the common lower bound for both ratios.
The inset shows the magnified figure of the ratio $r(x)$ for $k=\varphi ^{-1}$.
The lower and upper bounds are given by Eqs.~(\ref{eqn:t/xphi1}) and (\ref{eqn:t/xphi2}), respectively.
}
\end{figure}

\subsection{Asymptotic temporal behavior of the displacement $x(t)$}
The estimations given by Eqs.~(\ref{eqn:t/xf(x)}) and (\ref{eqn:f(x)}) indicate the fact that 
the ratio $t/x(t)$ has an intricate time dependence.
Using Eq.~(\ref{eqn:t/xf(x)}) recursively, the displacement $x(t)$ can be expressed as
\begin{equation}
x(t) \simeq \frac{t}{\alpha \ln
\displaystyle \frac{t}{\mathstrut \alpha \ln
\displaystyle \frac{t}{\mathstrut \alpha \ln
\displaystyle \frac{t}{\mathstrut \alpha \ln
\displaystyle \frac{t}{\mathstrut \ddots
}+f(x)}+f(x)}+f(x)}+f(x)}. \nonumber
\end{equation}
Provided that $\alpha \ln x \gg f(x),$ {\it i.e.},
$\alpha \ln Q_i \gg \frac{\theta _{i+1}}{\pi}+\frac{2\ln 2+1}{\pi}+C_1(k)$, is satisfied,
Eq.~(\ref{eqn:t/xf(x)}) is given by the $t\simeq \alpha x\ln x$, which can be solved.
Considering that the inverse function of $y=We^W$, the Lambert W function,
is written as $W(y)=-\log _e(\cdots \log _{(e^{-y})}\log _{(e^{-y})}(1/e))$ for $W\geq 1$ and $y\geq e$ \cite{rf:Yunh2001},
we have
\begin{equation}
x(t)\simeq \frac{1}{\cdots \mathfrak{L} _{t/\alpha} \circ \mathfrak{L} _{t/\alpha} \circ \mathfrak{L} _{t/\alpha} (1/e)} =\frac{1}{\mathfrak{L} _{t/\alpha} ^{\infty}(1/e)},
\label{eqn:continuedln}
\end{equation}
where the operater $\mathfrak{L}_{y}(u)$ is defined by $\mathfrak{L} _{y}(u)=\log _{(e^{-y})}u=\frac{1}{y}\ln \frac{1}{u}$.

The form of the nested logarithm is not trivial.
For $k=\varphi ^{-1}$, Eq.~(\ref{eqn:t/xphi1}), which represents the lower bound of $t/x$, can be solved as 
$x(t)=1/[ \beta \mathfrak{L} _{\beta t/\alpha} ^{\infty}(1/e)]$ with $\beta =e^{f_L/\alpha}$.
Here we set the $n$-th approximation of the lower bound, $x^{(n)}(t)=1/[ \beta \mathfrak{L} _{\beta t/\alpha} ^{n}(1/e)]$. 
Figure~5 shows the ratio $t/x$ for $k=\varphi ^{-1}$ as a function of $t$
with the first five approximations $t/x^{(n)}(t)\,\,(n=1,2,\cdots ,5)$ :
\begin{eqnarray}
\frac{t}{x^{(1)}}& = &  \alpha , \nonumber \\
\frac{t}{x^{(2)}}& = &  \alpha \ln \frac{t}{\alpha} +f_{L}, \nonumber \\
\frac{t}{x^{(3)}}& = & \alpha \ln
\displaystyle \frac{t}{\mathstrut \alpha \ln 
\displaystyle \frac{t}{\alpha}
+f_L}+f_L, \nonumber \\
\frac{t}{x^{(4)}}& = & \alpha \ln
\displaystyle \frac{t}{\mathstrut \alpha \ln
\displaystyle \frac{t}{\mathstrut \alpha \ln 
\displaystyle \frac{t}{\alpha}
+f_L}+f_L}+f_L,\nonumber 
\end{eqnarray}
and so on.
A higher-order logarithmic correction brings about better agreement with the numerical results.
The approximation functions $\{x^{(n)}(t)\}$ converge by oscillating to
a unique function $x^{(\infty )}(t)$ for large $t$,
as the order $n$ of logarithmic correction increases.
\begin{figure}[]%{6.6cm}   % l: LEFT, 6.6cm: WIDTH  
\includegraphics[scale=0.60, angle=-90]{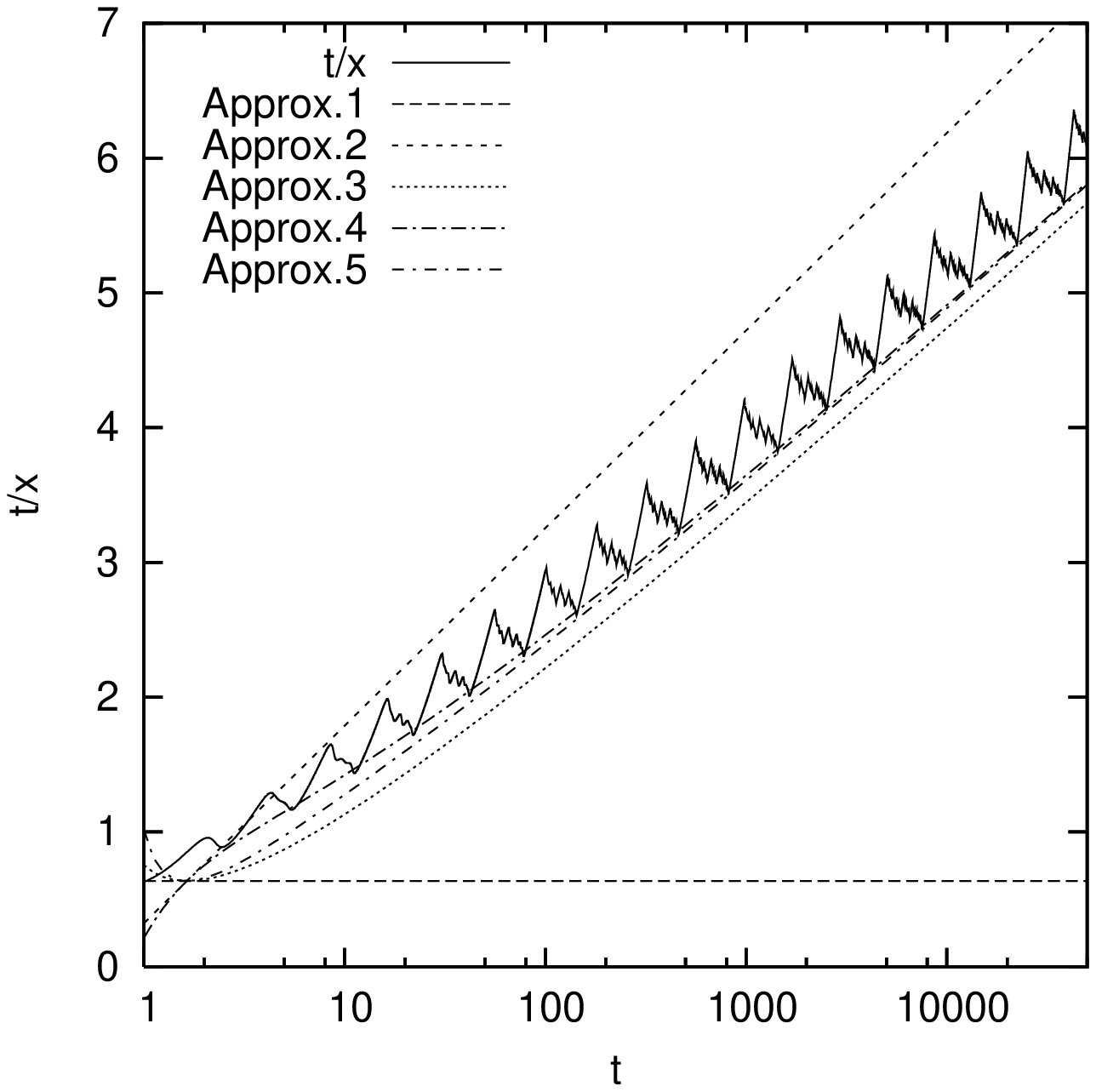}
\caption{Ratio $t/x$ as a function of $t$ (solid line) for $k=\varphi ^{-1}$.
The other lines are approximated curves $t/x^{(n)}(t)\,\,(n=1,2,\cdots ,5)$
for the lower bound.
}
\end{figure}

\section{Summary and discussion}
We have studied nonchaotic stagnant motion in the MQPGS.
It was shown that the density function of residence times
obeys the inverse-square law, independent of the parameter $k$. 
The spatial configuration of residence times, however, is sensitive to $k$.
The renewal of the largest residence time occurs
when the displacement $x(t)$ have passed through the cells $I_{Q_i}$, i.e.,
by the stagnation points $x\approx Q_i$,
which are denominators of the convergent sequence of parameter $k$.
Finally, the asymptotic behavior was given in the form of a nested logarithm. 

It is meaningful to compare the MQPGS
to intermittent chaotic systems \cite{rf:Mann1980,rf:Geis1984,rf:Gasp1988}.
In the PM system mentioned in Sec.~I,
the residence time $T$ of a laminar phase, starting at a reinjection point $x_{in}$
scales as $T\propto x_{in}^{-(z-1)}$ as $x_{in} \to 0$.
Since the probability density function (PDF) $P_{in}(x_{in})$ of the reinjection points varies only slowly with $x_{in}$,
{\it i.e.},  $P_{in}(x_{in}\to 0)\approx \mbox{const.}$, the PDF of residence times $P_{PM}(T)$ follows
an inverse-power law $P_{PM}(T)\propto T^{-\frac{z}{z-1}}$.
In particular, $P_{PM}(T)\propto T^{-2}$ for $z=2$,
as in the case of Eq.~(\ref{eqn:densitydist}).

Confining our discussion to the Pomeau-Manneville intermittency,
the following two factors are essential for the appearance of intermittency:
Local slow dynamics near the unstable periodic point, called intermissions, 
and a certain randomness in the seeds of residence time fluctuations.
The reinjection points $x_{in}$ are considered as the seeds in the PM system, 
and the parameters $\varepsilon _n$ are considered as the seeds in MQPGS.
For the two essential factors, the local mechanisms generating intermissions are almost the same in the two systems.
However, the seeds of residence time fluctuations have different properties in the systems. 
The reinjection points $x_{in}$ are random in the sense that
they are not only ergodic but also mixing (i.e., the correlation vanishes rapidly).
On the other hand, the parameters $\varepsilon _n$ have regularity,
since they are only ergodic and not mixing.

This difference is reflected in the asymptotic behavior, 
especially in the relaxation behavior of the long-time average.
In the PM system, the renewal rate $H(n)/n$  of chaotic bursts
converges as $1/\ln n$ for $z=2$,
which is derived based on the assumption that 
the successive residence times can be considered as independent random variables
\cite{rf:Mann1980,rf:Geis1984,rf:Gasp1988}.
In the MQPGS, however, the occurrence rate of rapid moving phase,
which coincides with $x(t)/t$, converges in the form of a nested logarithm.
The peculiar formula reflects the quasiperiodic correlation in $\varepsilon _n$.
We also note that the log-periodicity of the asymptotic behavior also appears
in quasichaotic systems, which exhibit weakly mixing dynamics
and have a zero Lyapunov exponent \cite{rf:Zasl2002}.
We believe that the MQPGS affords another model for investigating complex phenomena,
including slow relaxation, as well as nonstationarity,
from the viewpoint of regular system.

In Sec.~III we mentioned the structural instability of this system.
The stagnation points that generate the largest residence time are structurally unstable,
at which trajectories with slightly different parameter values can separate.
This instability depends on the number-theoretic properties of $k$.
We will investigate this aspect in a forthcoming paper \cite{rf:Mitslife}.

\section*{Acknowledgements}
The author thanks Professor Y.~Aizawa for valuable discussions and a great deal of encouragement, 
Dr. T.~Akimoto and Dr. T.~Miyaguchi for useful comments on the manuscript,
and Mr. S.~Shinkai for a lecture on the Lambert W function.
This work has been supported by a grant to the 21st-Century COE program, 
``Holistic Research and Education Center for Physics of Self-Organization Systems'',
at Waseda University from the Ministry of Education, Culture, Sports, Science, and Technology (MEXT), Japan.

\appendix	
\section{Derivation of Eq.~(\ref{eqn:eps})}
In the theory of continued fraction, the following formulas are known: \cite{rf:Rock1992}\\
(1) $P_i$'s and $Q_i$'s have the property that 
\begin{equation}
P_{i+1}Q_i-P_iQ_{i+1}=(-1)^i\,\,\,\,(i\geq -2).
\label{eqn:assist}
\end{equation}
(2) The parameter $k$ is written in another form,
\begin{equation}
k=\frac{\theta _{i+1}P_i+P_{i-1}}{\theta _{i+1}Q_i+Q_{i-1}}\,\,\,\,(i\geq -1).
\label{eqn:anotherform}
\end{equation}
Using Eqs.~(\ref{eqn:recursive}), (\ref{eqn:assist}), and (\ref{eqn:anotherform}),
the difference between the value of $k$ and its $i$-th convergent is given by 
\begin{equation}
k-\frac{P_i}{Q_i}=\frac{(-1)^i}{(\theta_{i+1} Q_i+Q_{i-1})Q_i}.\label{eqn:difference}
\end{equation}
Thus, 
$$\left| kQ_i-P_i\right| =\frac{1}{\theta_{i+1} Q_i+Q_{i-1}}.$$
Since the sequence $\{Q_i\}$ increases monotonically for $i$, and $\theta _i> 1$,
$$\left| kQ_i-P_i\right| < \left| kQ_1-P_1\right| =\frac{1}{\theta_{2}Q_1+Q_0}=\frac{1}{\theta_{2}q_1+1}<\frac{1}{2}.$$
The inequality $\left| kQ_i-P_i\right| < 1/2$ assures that $P_i$ is the nearest integer to $kQ_i$.
As a result, $\varepsilon _{Q_i}=\left| kQ_i-P_i\right| $ holds.

\section{Proof of theorem~1}
This is the proof with a slight modification of the original one given by Lagrange \cite{rf:Rock1992}. 
Consider the equation
$$kn-m=\alpha (kQ_{i+1}-P_{i+1})+\beta (kQ_{i}-P_{i}).$$
Separating the coefficient of $k$ from the constant terms,
we obtain two equations with two unknowns:
\begin{equation}
n=\alpha Q_{i+1}+\beta Q_i,\,\,m=\alpha P_{i+1}+\beta P_i \label{eqn:unknown}
\end{equation}
with a determinant given by Eq.~(\ref{eqn:assist}).
Since Eq.~(\ref{eqn:unknown}) is transformed as
$(-1)^i\alpha =mQ_i-nP_i,\,\,(-1)^i\beta =nP_{i+1}-mQ_{i+1},$
$\alpha$ and $\beta$ must be integers. In addition, $\alpha$ and $\beta$ cannot be zero, 
since $m/n$ is different from either $P_i/Q_i$ or $P_{i+1}/Q_{i+1}$. 
Furthermore, since $n\leq Q_{i+1}$, $\alpha$ and $\beta$ must have opposite signs.
Hence, since $kQ_{i+1}-P_{i+1}$ and $kQ_{i}-P_{i}$ also have opposite signs from Eq.~(\ref{eqn:difference}),
$$|kn-m|=|\alpha (kQ_{i+1}-P_{i+1})|+|\beta (kQ_{i}-P_{i})|>|kQ_{i}-P_{i}|.$$
Finally, we have $|kQ_{i}-P_{i}|>|kQ_{i+1}-P_{i+1}|$ from Eq.~(\ref{eqn:eps}). 
\hfill $\square$\\

\section{Proof of the inequality $\frac{T_n(\varepsilon _n)}{d\varepsilon _n}<0$}
Let $T(k,\,\varepsilon )$ denote the residence time $T_n(\varepsilon _n)$ defined by Eq.~(\ref{eqn:defT}) abbreviating $n$,
\begin{eqnarray*}
T(k,\varepsilon )&=&\int_{-\frac{1}{2}}^{\frac{1}{2}} \frac{dy}{1-\frac{1}{2}\cos (2\pi y)-\frac{1}{2}\cos (2\pi (\varepsilon +ky))} \\
&&\text{for}\,\,\,\,(k,\varepsilon )\in(0,1)\times (0,1/2).
\end{eqnarray*}
Then, the derivative of $T(k,\,\varepsilon )$ with respect to $\varepsilon$ is given by
\begin{eqnarray*}
\frac{dT(k,\varepsilon )}{d\varepsilon}&=&-4\pi \int_{-\frac{1}{2}}^{\frac{1}{2}}f(y)dy,\\
\text{where}\,\,\,\,f(y)&=&\frac{\sin(2\pi (\varepsilon +ky))}{(2-\cos (2\pi y)-\cos (2\pi (\varepsilon +ky)))^2}.
\end{eqnarray*}
To know the sign of the derivative, the parameter space is partitioned into following four regions:\\

\noindent {\bf Region~1}\,\,\,\,$\frac{k}{2}<\varepsilon <\frac{1-k}{2}$
$$f(y)>0\,\,\,\,\mbox{for}\,\,\,\,-\frac{1}{2}< y <\frac{1}{2}.$$
Thus, $\frac{dT}{d\varepsilon}<0.$\\
\begin{widetext}
\noindent {\bf Region~2}\,\,\,\,$\varepsilon \leq \min\left[ \frac{k}{2},\,\frac{1-k}{2} \right]$
\begin{eqnarray*}
\frac{dT(k,\varepsilon )}{d\varepsilon} &=& -4\pi \int_{-\frac{1}{2}}^{-\frac{\varepsilon}{k}}f(y)dy
                                            -4\pi \int_{-\frac{\varepsilon}{k}}^{\frac{1}{2}}f(y)dy,\\
                                        &=& -4\pi \int_{0}^{\frac{1}{2}-\frac{\varepsilon}{k}}f_1(z)dz
                                            -4\pi \int_{0}^{\frac{1}{2}+\frac{\varepsilon}{k}}f_2(z)dz,\\
                                        &=& -4\pi \int_{0}^{\frac{1}{2}-\frac{\varepsilon}{k}}(f_1(z)+f_2(z))dz
                                            -4\pi \int_{\frac{1}{2}-\frac{\varepsilon}{k}}^{\frac{1}{2}+\frac{\varepsilon}{k}}f_2(z)dz,
\end{eqnarray*}
where 
$f_1(z)=
-\frac{\sin(2\pi kz)}{(2-\cos (2\pi (z+\frac{\varepsilon}{k}))-\cos (2\pi kz))^2}<0,$ 
and 
$f_2(z)=\frac{\sin(2\pi kz)}{(2-\cos (2\pi (z-\frac{\varepsilon}{k}))-\cos (2\pi kz))^2}>0$ 
in each domain of integration without endpoints.
In addition,
\begin{equation}
f_1(z)+f_2(z)=
\frac{4\sin (2\pi z)\sin (2\pi kz)\sin (2\pi \frac{\varepsilon}{k})(2- \cos (2\pi kz)-\cos (2\pi z)\cos (2\pi \frac{\varepsilon}{k}))}
{(2-\cos (2\pi (z+\frac{\varepsilon}{k}))-\cos (2\pi kz))^2(2-\cos (2\pi (z-\frac{\varepsilon}{k}))-\cos (2\pi kz))^2}>0.\nonumber
\end{equation}
Hence, $\frac{dT}{d\varepsilon}<0.$\\

\noindent {\bf Region~3}\,\,\,\,$\varepsilon \geq \max \left[ \frac{k}{2},\,\frac{1-k}{2} \right]$
\begin{eqnarray*}
\frac{dT(k,\varepsilon )}{d\varepsilon} &=& -4\pi \int_{-\frac{1}{2}}^{\frac{1/2-\varepsilon}{k}}f(y)dy
                                            -4\pi \int_{\frac{1/2-\varepsilon}{k}}^{\frac{1}{2}}f(y)dy,\\
                                        &=& -4\pi \int_{0}^{\frac{1}{2}+\frac{1/2-\varepsilon}{k}}f_3(z)dz
                                            -4\pi \int_{0}^{\frac{1}{2}-\frac{1/2-\varepsilon}{k}}f_4(z)dz,\\
                                        &=& -4\pi \int_{0}^{\frac{1}{2}-\frac{1/2-\varepsilon}{k}}(f_3(z)+f_4(z))dz
                                            -4\pi \int_{\frac{1}{2}-\frac{1/2-\varepsilon}{k}}^{\frac{1}{2}-\frac{1/2-\varepsilon}{k}}f_3(z)dz,
\end{eqnarray*}
where 
$f_3(z)=\frac{\sin(2\pi kz)}{(2-\cos (2\pi (z-\frac{1/2-\varepsilon}{k}))+\cos (2\pi kz))^2}>0,$ 
and 
$f_4(z)=-\frac{\sin(2\pi kz)}{(2-\cos (2\pi (z+\frac{1/2-\varepsilon}{k}))+\cos (2\pi kz))^2}<0$ 
in each domain of integration without endpoints.
In addition,
\begin{equation}
f_3(z)+f_4(z)=
\frac{4\sin (2\pi z)\sin (2\pi kz)\sin (2\pi \frac{1/2-\varepsilon}{k})(2- \cos (2\pi kz)-\cos (2\pi z)\cos (2\pi \frac{1/2-\varepsilon}{k}))}
{(2-\cos (2\pi (z-\frac{1/2-\varepsilon}{k}))+\cos (2\pi kz))^2(2-\cos (2\pi (z+\frac{1/2-\varepsilon}{k}))+\cos (2\pi kz))^2}>0.
\end{equation}
Hence, $\frac{dT}{d\varepsilon}<0.$\\

\noindent {\bf Region~4}\,\,\,\,$\frac{1-k}{2}<\varepsilon <\frac{k}{2}$
\begin{eqnarray*}
\frac{dT(k,\varepsilon )}{d\varepsilon} &=& -4\pi \int_{-\frac{1}{2}}^{-\frac{\varepsilon}{k}}f(y)dy
                                            -4\pi \int_{-\frac{\varepsilon}{k}}^{\frac{1/2-\varepsilon}{k}}f(y)dy
                                            -4\pi \int_{\frac{1/2-\varepsilon}{k}}^{\frac{1}{2}}f(y)dy,\\
                                        &=& -4\pi \int_{0}^{\frac{1}{2}-\frac{\varepsilon}{k}}(f_1(z)+f_2(z))dz 
                                            -4\pi \int_{0}^{\frac{1}{2}-\frac{1/2-\varepsilon}{k}}(f_3(z)+f_4(z))dz
                                            -4\pi \int_{\frac{1}{2}-\frac{2\varepsilon}{k}}^{\frac{1-2\varepsilon}{k}-\frac{1}{2}}f(y)dy.
\end{eqnarray*}
\end{widetext}
The first and the second terms are negative, as proven in Regions~2 and 3, respectively.
The last term is negative since $f(y)>0$ in $\left[ \frac{1}{2}-\frac{2\varepsilon}{k},\,\,\frac{1-2\varepsilon}{k}-\frac{1}{2}\right].$ 
Thus, $\frac{dT}{d\varepsilon}<0.$

As a result, the monotonically decreasing property is proven. \hfill $\square$\\

\section{Proof of Eq.~(\ref{eqn:t/xmidineq})}
For a large $Q_i$, Eq.~(\ref{eqn:t/xmidineq}) is equivalent to the following inequalities:
\begin{eqnarray}
\frac{1}{Q_i}\sum _{j=1}^{Q_i}T_j \gtrsim 
\frac{1}{n}\sum _{j=1}^{n}T_j \gtrsim \frac{1}{Q_{i+1}-1}\sum _{j=1}^{Q_{i+1}-1}T_j \label{eqn:sum} \\
(n=Q_i,\,Q_i+1,\,\cdots ,Q_{i+1}-1),\nonumber
\end{eqnarray}

\noindent {\bf Proof of the first inequality}\\
The first inequality is equivalent to the following inequality:
\begin{eqnarray}
\frac{1}{Q_i}\sum _{j=1}^{Q_i}T_j \gtrsim
\frac{1}{n-Q_i}\sum _{j=Q_i+1}^{n}T_j \label{eqn:sumQi} \\
(n=Q_i+1,\,\cdots ,Q_{i+1}-1). \nonumber
\end{eqnarray}
Assume that the dynamics of $\{ \varepsilon _j\}_{j=Q_i+1}^{n}$ is approximately ergodic 
over the interval $[\varepsilon ^{\min},\,\varepsilon ^{\max}]$,
where $\varepsilon ^{\min}(>\varepsilon _{Q_i})$ and $\varepsilon ^{\max}$ are 
the minimum and maximum of $\varepsilon _j\,\,(j=Q_i+1,\,\cdots ,\,n)$, respectively.
Then, the summation of the right-hand side of Eq.~(\ref{eqn:sumQi}) can be
replaced with the average calculated by the density function $P(T)$ defined by Eq.~(\ref{eqn:densitydist}) as follows:
\begin{equation}
\frac{1}{n-Q_i}\sum _{j=Q_i+1}^{n}T_j = \int^{T^{\max}}_{T^{\min}}TP(T)dT < \int^{T_{Q_i}}_{T^{\min}}TP(T)dT, \label{eqn:point}
\end{equation}
where the last inequality comes from $T^{\max}<T_{Q_i}$, due to Corollary~1.
The last integral in Eq.~(\ref{eqn:point}) is given by
\begin{equation}
\int^{T_{Q_i}}_{T^{\min}}TP(T)dT
\simeq  \frac{2}{\pi}\ln (Q_{i}+Q_{i-1}/\theta _{i+1})+\frac{2}{\pi}\ln \theta _{i+1}+C_1(k).
\label{eqn:D4}
\end{equation}
Using Eq.~(\ref{eqn:D4}) and Eq.~(\ref{eqn:t/xQ+1/2}) for the left-right hand side of Eq.~(\ref{eqn:sumQi}), we find
\begin{eqnarray*}
\frac{1}{Q_i}\sum _{j=1}^{Q_i}T_j-\int^{T_{Q_i}}_{T^{\min}}TP(T)dT
\simeq  \frac{(\theta _{i+1}-2\ln \theta _{i+1})Q_{i}+Q_{i-1}}{\pi Q_i}>0,
\end{eqnarray*}
where $\theta -2\ln \theta >0$ is used.
Therefore, Eq.~(\ref{eqn:sumQi}), {\it i.e.}, the first inequality holds. \\

\noindent {\bf Proof of the second inequality}\\
The middle term in Eq.~(\ref{eqn:sum}) can be written as
$$\frac{1}{n}\sum _{j=1}^{n}T_j = \frac{1}{n} \left[ \sum _{j=1}^{Q_{i+1}-1}T_j-\sum _{j=n+1}^{Q_{i+1}-1}T_j\right].$$
Since the summation $\sum _{j=n+1}^{Q_{i+1}-1}T_j$ also does not have any term larger than $T_{Q_i}$, due to Corollary~1,
it is estimated as $$\sum _{j=n+1}^{Q_{i+1}-1}T_j\lesssim (Q_{i+1}-1-n)\int^{T_{Q_i}}_{T^{\min}}TP(T)dT.$$
where the ergodicity of $\{\varepsilon _j\}_{j=n+1}^{Q_{i+1}-1}$ is assumed again.
Hence,
\begin{widetext}
\begin{eqnarray*}
\frac{1}{n}\sum _{j=1}^{n}T_j &\gtrsim & \frac{1}{n} \Biggl[ (Q_{i+1}-1)\int^{T_{Q_i}}_{T^{\min}}TP(T)dT-(Q_{i+1}-1-n)\int^{T_{Q_i}}_{T^{\min}}TP(T)dT \Biggr] \\
&=& \int^{T_{Q_i}}_{T^{\min}}TP(T)dT \simeq \frac{1}{Q_{i+1}-1}\sum _{j=1}^{Q_{i+1}-1}T_j.
\end{eqnarray*}
\end{widetext}
Thus, the second inequality holds. \hfill $\square$\\

%\newpage %Just because of unusual number of tables stacked at end

\end{document}